\pdfoutput=1
\documentclass[letterpaper,english,aps,letter,superscriptaddress,nofootinbib,floatfix,pre,preprint]{revtex4}
\usepackage[T1]{fontenc}
\usepackage[latin9]{inputenc}
\usepackage{amsbsy}
\usepackage{amstext}
\usepackage{graphicx}
\usepackage{amssymb}

\makeatletter
\@ifundefined{textcolor}{}
{%
 \definecolor{BLACK}{gray}{0}
 \definecolor{WHITE}{gray}{1}
 \definecolor{RED}{rgb}{1,0,0}
 \definecolor{GREEN}{rgb}{0,1,0}
 \definecolor{BLUE}{rgb}{0,0,1}
 \definecolor{CYAN}{cmyk}{1,0,0,0}
 \definecolor{MAGENTA}{cmyk}{0,1,0,0}
 \definecolor{YELLOW}{cmyk}{0,0,1,0}
 }

\usepackage{ae,aecompl}
\renewcommand{\citet}[1]{\cite{#1}}

\usepackage{babel}

\makeatother

\usepackage{babel}

\begin{document}

\title{First-Passage Kinetic Monte Carlo method}

\author{Tomas Oppelstrup}

\affiliation{Lawrence Livermore National Laboratory, Livermore, California 94551,
USA}

\affiliation{Royal Institute of Technology (KTH), Stockholm S-10044, Sweden}

\author{Vasily V. Bulatov}

\affiliation{Lawrence Livermore National Laboratory, Livermore, California 94551,
USA}

\author{Aleksandar Donev}

\affiliation{Lawrence Livermore National Laboratory, Livermore, California 94551,
USA}

\affiliation{Center for Computational Science and Engineering, Lawrence Berkeley
National Laboratory, Berkeley, CA, 94720}

\author{Malvin H. Kalos}

\affiliation{Lawrence Livermore National Laboratory, Livermore, California 94551,
USA}

\author{George H. Gilmer}

\affiliation{Lawrence Livermore National Laboratory, Livermore, California 94551,
USA}

\author{Babak Sadigh}

\affiliation{Lawrence Livermore National Laboratory, Livermore, California 94551,
USA}
\begin{abstract}
We present a new efficient method for Monte Carlo simulations of diffusion-reaction
processes. First introduced by us in {[}\emph{Phys. Rev. Lett.}, 97:230602,
2006{]}, the new algorithm skips the traditional small diffusion hops
and propagates the diffusing particles over long distances through
a sequence of super-hops, one particle at a time. By partitioning
the simulation space into non-overlapping protecting domains each
containing only one or two particles, the algorithm factorizes the
$N$-body problem of collisions among multiple Brownian particles
into a set of much simpler single-body and two-body problems. Efficient
propagation of particles inside their protective domains is enabled
through the use of time-dependent Green's functions (propagators)
obtained as solutions for the first-passage statistics of random walks.
The resulting Monte Carlo algorithm is event-driven and asynchronous;
each Brownian particle propagates inside its own protective domain
and on its own time clock. The algorithm reproduces the statistics
of the underlying Monte-Carlo model exactly. Extensive numerical examples
demonstrate that for an important class of diffusion-reaction models
the new algorithm is efficient at low particle densities, where other
existing algorithms slow down severely.
\end{abstract}
\maketitle
\global\long\def\Cross#1{\left|\mathbf{#1}\right|_{\times}}
\global\long\def\CrossL#1{\left|\mathbf{#1}\right|_{\times}^{L}}
\global\long\def\CrossR#1{\left|\mathbf{#1}\right|_{\times}^{R}}
\global\long\def\CrossS#1{\left|\mathbf{#1}\right|_{\boxtimes}}

\global\long\def\V#1{\mathbf{#1}}
\global\long\def\M#1{\mathbf{#1}}
\global\long\def\D#1{\Delta#1}

\global\long\def\sV#1{\boldsymbol{#1}}
\global\long\def\sM#1{\boldsymbol{#1}}

\global\long\def\grad{\boldsymbol{\nabla}}
\global\long\def\eij{\left\{  i,j\right\}  }
\global\long\def\celsius{^{\circ}C}

\section{Introduction}

Models in which the overall dynamics is represented by random walks
are widely applied in science, engineering, medicine and finance.
Probably the simplest example of a random walk is a sequence of steps
taken randomly in two directions -- left or right -- along a line
in one dimension. The object whose displacements follow such a sequence
is referred to as a random walker or, simply, a walker. Of particular
interest are diffusion-reaction systems in which multiple walkers
walk simultaneously and independently and some significant events
take place when two or more walkers find each other in space, or collide.
Examples include formation and growth of aggregates of colloidal particles
in suspensions, kinetics of aerosols in meteorology, diffusive phase
transformations in solids \citet{Coarsening_AtomisticMeanField},
surface diffusion during crystal growth from vapor \citet{KMC_SurfaceReactions,LG_KMC_EpitaxialGrowth},
defect evolution in solids \citet{BIGMAC,LAKIMOCA_OKMC}, multi-particle
diffusion-limited aggregation in physics, diffusion-controlled reactions
in chemistry and biochemistry \citet{ParticleBased_CellModeling,GFRD_KMC,DiffusionReaction_Plimpton},
population dynamics, quantum physics \citet{KLV_PRE}, and risk assessment
and pricing models in finance to name a few. Numerical simulations
of such processes often utilize various flavors of the Monte Carlo
method.

Kinetic Monte Carlo (KMC) is a simple and robust computational approach
for simulations of systems evolving through random walks. Mathematically,
KMC derives from the theory of Markov processes in which the model
evolves from state to state through a sequence of stochastic transitions
whose rates depend on the current state alone. Random walks are typically
simulated as sequences of hops, either from one lattice site to a
neighboring one for discrete walks, or through finite displacements
for continuum walks. When the system dynamics is defined by collisions
among the walkers, the hops themselves are trivial changes of the
system's state while significant events take place only when the walkers
collide. A serious computational bottleneck is presented for the KMC
method by situations when the density of walkers is low. Consider
a system of randomly distributed walkers. It takes on average $\propto L^{3}$
hops for a walker to collide with another in $3d$ space (Here, $L$
is the average spacing between the walkers expressed in the units
of the lattice spacing or, in the continuum case, in the units of
particle diameter). When $L$ is large, it can take a great number
of KMC cycles to evolve the model to a meaningful event, a collision.
This is a serious drawback limiting applicability of the KMC method
to diffusion-reaction simulations.

Several attempts have been made so far to overcome this notorious
inefficiency in KMC simulations. In \citet{GFRD_KMC}, the equivalence
between continuous random walks and diffusion is exploited by using
the fundamental solution for the single particle diffusion to propagate
the walkers over large distances. The JERK method \citet{JERK} uses
a known solution for the statistics of binary collisions between two
diffusing particles to decide which of the $N(N-1)/2$ pairs of walkers
should collide over the next time step %
\footnote{It was in fact the JERK method and its applications to modeling irradiated
materials that inspired the key idea of the new algorithm presented
in this paper.%
}. These and similar methods achieve improvements in the efficiency
of KMC simulations but at a cost of their accuracy. The fundamental
difficulty that none of the mentioned methods can fully address is
that statistics of collisions in the system of $N$ walkers is an
$N$-body problem. That is, the probability of collisions between,
say, walkers 1 and 2, depends on all other $N-2$ walkers in the system.
It is only in the limit of very small hops (in \citet{GFRD_KMC})
or vanishing time steps (in \citet{JERK}) that such approximate methods
become asymptotically exact. Unfortunately, in this same limit the
mentioned methods lose their numerical efficiency.

Here we present a novel approach for KMC simulations that is both
efficient and exact for a wide class of models involving collisions
among multiple Brownian particles, as first proposed in Ref. \citet{FPKMC_PRL}.
Based on exact solutions for the first passage statistics of random
walks, the new method is referred to as First Passage Kinetic Monte
Carlo (FPKMC) in the following. In the new algorithm, rather than
propagating the particles to collisions by small diffusional hops,
the particles are propagated over long distances while each walker
(particle) is protected (separated from interference by other walkers)
within its own spatial region. The $N$ regions are non-overlapping
and partition the space into disjoint spatial domains in which the
enclosed walkers are propagated individually. The use of first-passage
statistics for walker propagation permits an elegant factorization
of the $N$-body problem into a product of $N$ single-body problems.
Efficient implementation of the new method leads to an asynchronous
event-driven algorithm \citet{AED_Review} in which every walker propagates
within its personal space and from its own time origin. The resulting
speedup is most impressive when the density of diffusing particles
is low and particle collisions are rare. 

In this paper we introduce the basic theory of the FPKMC method and
present a few simple but representative simulation tests on the method's
accuracy and efficiency. The paper is organized as follows. The next
section introduces the basic ideas behind the new method using one-dimensional
continuous random walks as a simple example. Section 3 describes the
overall algorithm. In section 4 we focus on propagators, i.e., elementary
solutions for first-passage statistics required for efficient propagation
of multiple walkers to collisions, and describe extensions of the
FPKMC algorithm to higher dimensions. Section 5 presents several computational
experiments validating the new method's accuracy and efficiency. The
results are summarized in Section 6. Appendix A describes a rejection
sampling procedure used in the FPKMC algorithm and Appendix B contains
a concise derivation of the propagators.

\section{First passage propagation in $1d$}

In this section we introduce the FPKMC algorithm using the continuous
limit of a random walk, i.e., a one-dimensional ($1d$) continuous
diffusion (Weiner) process, i.e., , as an illustrative example. An
extension to diffusion in dimensions higher than one will be described
in section \ref{sub:higher_d}.

With appropriate modifications, the FPKMC algorithm is also applicable
to simulations of other types of Markov random walks, such as jump
Markov processes on the continuum or jump Markov processes on a lattice.
The definitions and the algorithms to be presented here remain essentially
the same for discrete walks, but for discrete-valued space $x$ and/or
time $t$ the integrals appearing in the discussion below correspond
to sums over appropriate discrete values. We defer to a future publication
algorithmic details specific to discrete random walks.

\subsection{Single walker}

To define the probability distributions to be employed in the FPKMC
algorithm, let us first consider a single continuous random walk in
one dimension ($1d$). Let $x_{0}$ and $t_{0}=0$ be the position
and time origins of the walk and $a$ be some other (barrier) position
on the line $-\infty<x<\infty$. Through a sequence of random displacements
the walker can at some future time reach the barrier position $a$
for the first time. Similarly, for a closed interval $[a,b]$ such
that $a<x_{0}<b$ , a first passage event occurs when the walker reaches
either one of two barriers $a$ or $b$. The theory of first passage
processes \citet{FirstPassage_Redner} concerns itself with finding
the probability that the walker will reach one of two barriers for
the first time within time interval $[t,t+dt)$. The relevant statistical
distribution is the probability density $c(x_{0,}x,t)$ to find the
walker \emph{surviving} at time $t$ (having not reached either end
of $[a,b]$) within infinitesimal interval $[x,x+dx]$ inside $[a,b]$.
By its definition, the integral \begin{equation}
S(x_{0},t)=\intop_{a}^{b}dx'c(x_{0,}x',t)\label{eq:S_x0_t_def}\end{equation}
 is the total probability for the walker to survive by time $t$ regardless
of its end position $x$. The splitting probability $j(a,t)$ is defined
as the conditional probability that, given that the first-passage
event occurs at time $t$, the walker reaches barrier $a$ rather
than $b$. For a random walk in $1d$, $j(a,t)+j(b,t)=1$ at all times.
When the walk origin $x_{0}$ is exactly in the center of interval
$[a,b]$, the splitting probabilities are equal $j(a,t)=j(b,t)=\frac{1}{2}$
and independent of the first passage time $t$. Finally, the no-passage
(NP) probability distribution function (PDF) is defined as the conditional
probability to find the walker at position $x$ at time $t$, provided
the first-passage event has not yet occurred, \begin{equation}
g(x_{0},x,t)=\frac{c(x_{0},x,t)}{S(x_{0},t)}.\label{eq:g_x0_t_def}\end{equation}

We defer to section \ref{sec:The-propagators} the derivation of the
probability distributions introduced above. For now let us simply
assume that the functions $S(x_{0},t)$, $j(a,t)$ and $g(x_{o},x,t)$
are available and proceed to describe how they can be used to obtain
statistical samples of random walks in various situations.

First consider the statistics of continuous random walks on the line
$-\infty<x<\infty$ with no barriers. Assuming that the walks start
at position $x_{0}=0$ and time $t_{0}=0$, the PDF of walker positions
at time $t>0$ is given by the fundamental solution of the diffusion
equation \begin{equation}
c_{\infty}(x,t)=\frac{1}{\sqrt{4\pi Dt}}\exp(-\frac{x^{2}}{4Dt}),\label{eq:c_inf_gaussian}\end{equation}
where $D$ is the diffusion coefficient. The same statistics can be
obtained by randomly sampling from the first-passage (FP) and no-passage
(NP) distribution functions (propagators) as follows. Define an interval
of length $L_{1}$ centered on the initial walker position $x_{0}=0$.
Draw a random number $\xi$ uniformly distributed on $[0,1]$, henceforth
simply called a {}``random number'', and solve $S(L_{1},t_{1})=\xi$
to sample the exit time $t_{1}$ out of interval $[-\frac{L_{1}}{2},\frac{L_{1}}{2}]$
. If $t_{1}>t$, use the NP distribution $g_{L_{1}}(x,t)$ to sample
the walker position inside the interval. If $t_{1}<t$, use another
random number to sample at which end of interval $L_{1}$ the walker
exits at time $t_{1}$. Define a new interval of length $L_{2}$ centered
on the new walker position and sample a new time $t_{2}$ of first-passage
out of interval $L_{2}$ using the survival probability distribution
$S(L_{2},t)$. Continue until the sum of first-passage times $T_{k}=\sum_{i=1}^{k}t_{i}$
exceeds $t$. Use the NP propagator $g_{L_{k}}(x,t-T_{k-1})$ to sample
the end position of the walker. Proceeding in this manner, a random
sample of the walker position for any time $t$ is obtained through
a sequence of $k\geq0$ first-passage propagations ending in a single
no-passage propagation. Repeating such stochastic sampling sequences
many times, statistics of the end walker positions can be used to
reproduce the fundamental solution $c_{\infty}(x,t)$ to any desired
accuracy. The length $L$ of the propagation intervals defines how
many FP steps on average will be used to reach time $t$ but otherwise
has no bearing on the resulting statistics. 

The above example illustrates the use of NP and FP distributions for
sampling random walks on $-\infty<x<\infty$. The resulting samples
are statistically equivalent to the known distribution $c_{\infty}(x,t)$.
While not necessary in this particularly simple case, the same sampling
procedure based on the first-passage statistics can be effectively
employed in considerably more complex situations, such as the one
described below where the simple fundamental solution $c_{\infty}$
no longer applies.

\subsection{\label{sub:Multiple-walkers}Multiple walkers}

Consider now multiple objects moving randomly and simultaneously on
a properly defined space and time. The walkers are assumed to walk
independently of each other until two of them find themselves at a
distance equal to or smaller than some interaction radius $r$ (in
$1d$ the interaction radius can be set to zero). As was discussed
in the introduction, models of this kind represent a plethora of situations
of practical interest. For our discussion here it is not necessary
to define what specifically happens when the walkers reach the interaction
radius; let us just assume that collisions somehow affect propagation
statistics of two (or more) walkers involved in a collision.

The most straightforward numerical approach to modeling such systems
is to use random numbers to move the walkers over space by small hops,
one walker and one hop at a time, and checking after each such hop
if any of the walkers have collided. Although widely used, this simple
method is known to become less and less efficient with the decreasing
density of walkers \citet{DiffusionControlled_RRC}. The idea of the
method presented in this section is to circumvent the need for the
numerous small hops by using the solutions for the first passage statistics
of a single walker to efficiently bring the walkers to collisions.

Consider two simultaneous walks with the same time origin $t=0$ but
different position origins $x_{1}$ and $x_{2}$ in one dimension,
such that $-\infty<x_{1}<x_{2}<\infty$. Is it possible to obtain
statistics of collisions between two walkers using a sampling procedure
similar to the sequence of FP and NP propagations described in the
previous section? At a first glance, the answer should be negative
because, in principle, a collision between the walkers can occur at
any time (at least in the case of continuum diffusion), thus altering
the statistics of both walkers. Hence, the simple solutions for first
passage statistics of a single walker should no longer apply. Fortunately,
the trick of spatial \emph{protection} enables the use of single-walker
propagations.

\begin{figure}
\begin{centering}
\includegraphics[width=0.8\textwidth]{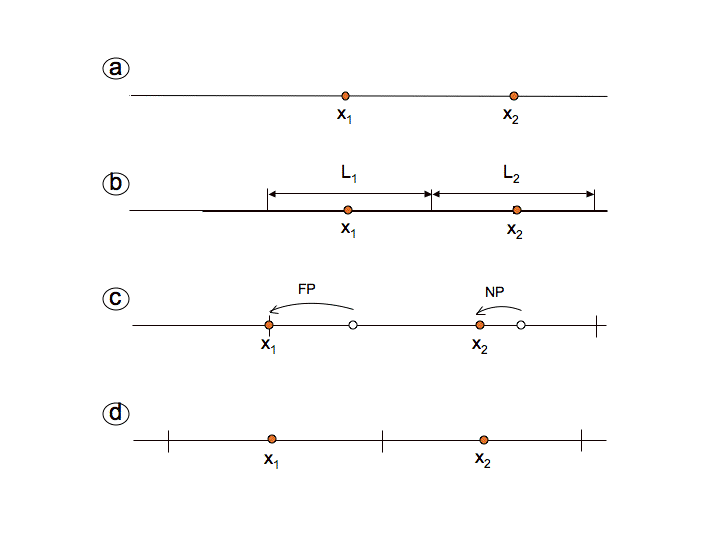}
\par\end{centering}

\caption{\label{fig:multiple} Simultaneous propagation of two walkers using
first-passage and no-passage distributions. The filled circles are
walker positions at different stages of propagation. (a) The walkers
are initially at positions $x_{1}$ and $x_{2}$ at time $t=0$. (b)
The walkers are protected by two non-overlapping intervals $L_{1}$
and $L_{2}$ centered on the walker positions. (c) Random samples
of two first passage times $t_{1}$ and $t_{2}$ are obtained and
compared. Because $t_{1}<t_{2}$, walker 1 is moved from its initial
position (open circle) to one of the ends of its protective segment.
At the same time, new position of walker 2 inside its segment $L_{2}$
is sampled from the NP distribution. Time advances to $t_{c}:=t_{1}$.
(d) A new propagation cycle begins by constructing non-overlapping
protective segments around new walker positions. Although after the
cycle illustrated by the (a)-(b)-(c) sequence the walkers find each
other slightly farther apart, the next cycle is just as likely to
bring them closer together. }

\end{figure}

Let us, at $t=0$, surround the walkers by two non-overlapping segments
$L_{1}$ and $L_{2}$ centered on the walk origins $x_{1}$ and $x_{2}$.
For example, make the ends of two segments coincide at the mid-point
between $x_{1}$ and $x_{2}$, as shown in Fig. \ref{fig:multiple}.
The key observation that enables the use of single-walker propagators
is that, for as long as both walks remain inside their segments, they
are protected from interactions with other walkers. Hence, up until
the time one of the walkers exits its protective segment, the statistics
of the two walks remains independent of each other and the single
walker propagators can be used. Let us now use the survival probabilities
$S(L_{1},t_{1})$ and $S(L_{2},t_{2})$ to randomly sample the first-passage
times $t_{1}$ and $t_{2}$ and find their minimum $t_{min}=\min\{t_{1},t_{2}\}$.
Say, $t_{min}=t_{1}$ which means that at time $t=t_{1}$ walker 1
reaches one of the ends of its protective segment $L_{1}$ while walker
2 remains inside its protective segment $L_{2}$. Let us randomly
select to which end of its protective segment walker 1 propagates
and advance the time clock by $t_{1}$, $t_{c}:=t_{1}$. Since walker
2 has not exited its protective segment $L_{2,}$, its new position
can be obtained by sampling from the NP distribution $g_{L_{2}}(x_{2,}t_{c}).$
Now the walkers find themselves in new positions $x_{1}$ and $x_{2}$
at time $t_{c}$, and the propagation cycle can be repeated: new protective
segments $L_{1}$ and $L_{2}$ are defined around the new walker positions,
two FP times are sampled and compared, the time clock is advanced
and new walker positions are sampled from appropriate FP and NP distributions.

Extension from two to $N$ walkers is straightforward. One starts
by defining non-overlapping protective segments ${L_{1},L_{2},...,L_{N}}$
centered on each walker and sampling first passage times for every
walker, $\{t_{1},t_{2},...,t_{N}\}$. For as long as all walkers remain
inside their protective segments, no walker can affect the statistics
of any other walker. Therefore, the use of single walker propagators
guarantees correct sampling of random walks at least until the next
scheduled propagation at time $t_{min}=\min\{t_{1},t_{2},...,t_{N}\}$.
At this time, walker $i$ with the shortest exit time $t_{i}=t_{min}$
is FP-propagated to one of the ends of its protective segment and
positions of all other $N-1$ walkers are sampled from appropriate
NP distributions.

The sampling procedure described above allows a seemingly small but
important modification: rather than canceling all exit times larger
than $t_{m}$ and NP-sampling new positions for the corresponding
$N-1$ walkers, \emph{all or almost all} of these $N-1$ walkers can
be left alone, protected inside their old segments and scheduled for
propagation at their previously sampled exit times. First, the exit
times sampled for all $N$ walkers are arranged in a priority queue
$t_{i}\leq t_{k}...\leq t_{m}$. Second, walker $i$ with the shortest
exit time is FP propagated to one of the ends of its protective segment
$L_{i}.$ The NP propagation is only necessary if and when the segment
end where walker $i$ exits is also shared by a neighboring protective
segment $L_{j}$. When this happens, independence of two affected
walks at later times is no longer assured because walker $i$ now
intrudes in the protective segment of walker $j$. The impasse is
resolved by sampling a new position for walker $j$ at time $t_{min}$
using the NP distribution. Now that current positions of walkers $i$
and $j$ are decided, the time advances to $t_{c}$ and walkers $i$
and $j$ are protected again in the space left available for them
by all other $N-2$ protective segments. Two new exit times are sampled
for walkers $i$ and $j$, added to the current time and inserted
in the queue. Preceding in this way, every cycle entails exactly one
FP propagation and at most one NP propagation, rather than $N-1$
NP propagations as in the algorithm proposed in Ref. \citet{KLV_PRE}. 

That it it is unnecessary to sample new exit times at $t=t_{min}$
for the walkers whose protective segments remain unaffected, follows
directly from the basic property of the random walk as a memory-less
stochastic process. In particular, for any $t_{1}<t$ and $x\in[a,b]$
the Chapman-Kolmogorov-Smoluchowski identity holds

\[
c(x_{0},x,t)=\intop_{a}^{b}dx'c(x_{0},x',t_{1})c(x',x,t-t_{1}).\]
Dividing both parts of the above equality by $S(x_{0},t_{1})=\intop_{a}^{b}dxc(x_{0},x,t)$
we obtain

\[
\frac{c(x_{0},x,t)}{S(x_{0},t_{1})}=\intop_{a}^{b}dx'g(x_{0},x',t_{1})c(x',x,t-t_{1}).\]
The expression on the left hand side is the probability density at
time $t$ of walks that started at $t=0$ and are known to have survived
at time $t_{1}$. The expression on the right hand side defines the
probability density of walks at time $t$ that have survived to time
$t_{1}$, when their positions $x'$ inside the interval were sampled
from the NP distribution $g(x_{0},x',t_{1})$, and the walk was then
restarted from the new position origin $x'$ and time origin $t_{1}$.
Integration of both sides of the above equality over $\intop_{a}^{b}dx$
yields the corresponding equality for the survival probabilities

\[
\frac{S(x_{0},t)}{S(x_{0},t_{1})}=\intop_{a}^{b}dx'g(x_{0},x',t_{1})S(x',t-t_{1}).\]
The expression on the right is the probability to survive at time
$t$ for a walk that has survived at $t_{1}$ and whose new position
$x'$ inside the interval was sampled from the NP distribution $g(x_{0},x',t_{1})$.
The expression on the left is the probability to survive at time $t$
for a walk that started at $t=0$ and survived at $t_{1.}$ The last
two equalities mean that for all walkers whose protective intervals
are unaffected by the FP propagation at $t_{1}$, there is no need
to sample new positions and new exit times because the resulting distributions
will be identical to the pre-sampled statistics. Thus, the two sampling
procedures - one used in Ref. \citet{KLV_PRE} and one proposed here
- are statistically equivalent. Obviously, the new procedure is much
preferred since the cost of its sampling cycle is not higher than
the cost of the queue update, i.e., $O(\log N)$, whereas the cost
of every sampling cycle in Ref. \citet{KLV_PRE} is $O(N)$. 

Just as in Ref. \citet{KLV_PRE}, in the new algorithm all $N$ walks
are initially protected and start from the same time origin $t_{0}=0$.
The walker with the shortest exit time $t_{min}$ is FP-propagated
and, perhaps, another neighboring walker is NP-propagated to new positions.
The global time clock advances to $t_{min}$ and the affected walkers
are protected by new segments. One or two new FP times are sampled,
added to the new global time and inserted in the time queue. Over
subsequent cycles, the time origins of the $N$ protected walkers
will gradually become desynchronized. Notwithstanding, statistical
independence of protected walkers is guaranteed up to the shortest
exit time in the current time queue. The resulting first-passage kinetic
Monte Carlo (FPKMC) algorithm is asynchronous: every walker propagates
within its personal space (protective segment) and from its own position
and time origins. Sooner or later, a series of FP and NP propagations
executed in this manner should bring a pair of walkers close to their
interaction radius.

As further discussed in section \ref{sec:FPKMCAlgorithm}, efficiency
of the FPKMC algorithm demands special treatment of colliding pairs
of particles in order to prevent the lengths of their protective segments
from shrinking to zero as the walkers approach each other. Namely,
by allowing two protective segments segments to overlap the possibility
of a collision between any two neighboring walkers can be included
as a possible propagation outcome. Such pair propagations entail sampling
from an appropriate Green's function (\emph{pair propagator}), as
explained in section \ref{sub:Pair-Propagators}. Similar considerations
apply to collisions between the walkers and the surfaces (e.g., absorbing
or reflective boundaries). Specifically, for a walker near a boundary
the segment $L_{1}$ can be made to touch the boundary of the domain
so that one of the possible propagation outcomes corresponds to a
collision (absorption or reflection) with the boundary.

Barring any inaccuracies in the single-particle or pair propagators,
the new algorithm is as exact as a Monte Carlo algorithm can be: for
any number of walkers $N$, the statistics of simultaneous random
walks with collisions is correctly reproduced in the limit of large
number of independent Monte Carlo simulations.

\section{\label{sec:FPKMCAlgorithm}The FPKMC algorithm }

Here we give a brief description of the algorithmic components necessary
for an FPKMC implementation.

First, one has to obtain first-passage (FP) and no-passage (NP) propagators.
The NP propagators are needed when a walker propagates right on or
close to the boundary of a neighboring protective segment. In such
a case, the new protective segment for the just propagated walker
will have very small (or even zero) length. Consequently, the new
time for FP propagation of the {}``squeezed'' walker is likely to
be so short that the same walker will be selected for the very next
propagation again. Because its protection is tightly constrained by
the protective segments of its neighbor walkers, the squeezed walker
would continue to perform a series of very short FP propagations resulting
in little and, eventually, no advance of the global clock. The solution
to this inefficiency of the \emph{Achilles and Turtle} type is to
identify which neighbor (or neighbors) limit the space available for
protection of the squeezed walker. Then, NP propagation of the constraining
inactive neighbor walker(s) typically results in more equitable partitioning
of space with the squeezed walker. This is achieved at the cost of
canceling the earlier scheduled FP propagation of the constraining
walker(s) and propagating it (them) using the NP propagator to the
current time. Proceeding in this manner, every Monte Carlo cycle entails
one FP propagation and, possibly, one or few NP propagation, while
all other $N-1$ or $N-2$ walkers stay inactive, scheduled for propagations
at their own times in the future. Calculation and efficient use of
FP and NP propagators are discussed in the next section and, in more
detail, in the Appendices.

In FPKMC, FP and NP propagations replace numerous short diffusive
hops. At the same time, much of the computational effort is shifted
to maintaining efficient space partitioning among the protective domains
of the walkers. It is useful to observe that for the FPKMC method
to work, space can be partitioned in an arbitrary manner for as long
as the protective domains remain non-overlapping. One can use this
freedom to simplify implementation and to maximize computational efficiency.
To minimize implementation effort, we use particularly simple protective
domains, i.e. centered segments (in $1d$) and centered hyper-cubes
(in dimensions $d\geq2$). 

Generally, one would want to partition space for maximum computational
efficiency, for example to maximize the expectation time of the next
FP propagation event. Optimal space partitioning for arbitrary positions
of the $N$ walkers can be accomplished in $O(N)$ operations - we
do it only sparingly, such as in the beginning of each simulation
run. During the run, the conditions of space sharing are inspected
only for the walkers that were just propagated, and their immediate
neighbors. Definition of optimal space partitioning depends on the
relative mobilities of the walkers. In this paper we only consider
models in which all $N$ walkers have exactly the same mobility properties,
i.e., the same diffusion coefficient for the case of continuum random
walks or the same hopping rates for the case of discrete walks. Cases
when some walkers are more mobile than others will be considered in
a future publication.

The FPKMC algorithm allows exact and efficient treatment of particle
collisions, by protecting and propagating groups of walkers, e.g.
pairs. As we describe in detail in Section \ref{sub:Pair-Propagators},
multi-particle propagators needed for such an extension are particularly
simple in $1d$ and for hypercube-shaped walkers for $d\geq2$. The
evolution of various diffusion-reaction models is defined by what
actually happens to colliding walkers - annihilation, coalescence,
reflection, etc. - and how the collisions are handled. The same details
will obviously affect the efficiency of FPKMC simulations but, in
our view, the ultimate purpose of the FPKMC method is to enable efficient
propagation of walkers to collisions whereas handling of collisions
events is outside of the method's main scope. Thus, FPKMC can be viewed
as a universal accelerator for particle diffusion or random walks
by which the particles or walkers are brought to or \emph{close to
collisions}. We leave it for the method's users to define collision
outcomes and to develop accurate and efficient methods for collision
handling. To keep it simple, in this paper we consider only annihilation
and coalescence reactions leaving more complicated collision scenarios
for future publications.

In summary, the FPKMC algorithm entails the following steps: 
\begin{enumerate}
\item Set the global time clock to zero. Construct non-overlapping protective
domains around all walkers - use individual protection for single
walkers and group protection for close pairs, as seems most efficient. 
\item Sample an exit time for each domain (in the case of protected pairs
this can mean a scheduled collision). Put the sampled event times
in an event queue (e.g., implemented as a heap), so that the shortest
time can be efficiently found. 
\item Find the shortest exit time and identify the corresponding walker
and domain. Sample the exit position for the selected walker. If the
new position corresponds to a collision, take appropriate action. 
\item Check if any of the existing protective domains are close to the new
position of the particle. If necessary to make more space available
for protection of the propagated particle, use no-passage propagators
to sample new locations for the particles in the neighboring domains. 
\item Construct new protective domains for all particles that changed their
positions in steps (3) or (4). 
\item Sample new event times for the particle(s) protected in step (5),
as in step (2). 
\item Insert the new event time(s) into the event queue. Go to step (3). 
\end{enumerate}

\section{\label{sec:The-propagators}The propagators}

The FPKMC algorithm relies on the first-passage (FP) and no-passage
(NP) propagators to skip the numerous small steps and to bring the
walkers to collisions. For the new algorithm to be efficient, Monte
Carlo sampling from these propagators should not entail significant
computational overhead. In this section, the elementary mathematical
theory behind the propagators is presented along with explicit propagator
formulas for the case of continuous diffusion in $1d$. We also discuss
two methods appropriate for Monte Carlo sampling from the FP and NP
propagators. Propagators suitable for efficient FPKMC simulations
of $N$ simultaneous random walkers on lattices will be given in future
publications.

\subsection{Continuous space - continuous time }

\begin{figure}
\begin{centering}
\includegraphics[width=0.5\textwidth]{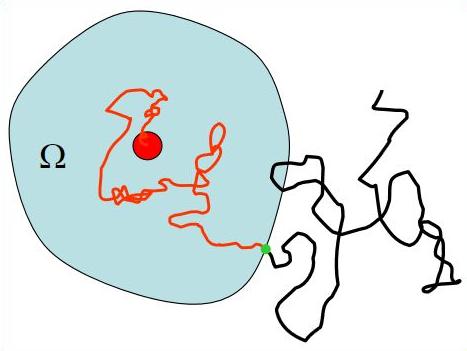} 
\par\end{centering}

\caption{\label{fig:omega}Consider a random walk and a domain $\Omega$ enclosing
the walk origin. The walk can be thought to consist of two parts:
the part shown in red that is entirely contained inside $\Omega$,
and the rest of the walk shown in black. }

\end{figure}

The problem we consider here is to find when and where a particle
performing a random walk (continuous diffusion process) exits a specified
domain. Figure \ref{fig:omega} shows a schematic of the setting.

Statistics of continuum random walks is equivalent to diffusion. Consider
a very large ensemble of non-interacting random walkers starting simultaneously
from the same origin. The concentration of walkers in this ensemble
is the solution to the diffusion equation with a delta function at
the walk origin as the initial condition

\begin{equation}
D\Delta c(\bar{x},t)=\frac{\partial c(\bar{x},t)}{\partial t},\quad c(\partial\Omega,t)=0,\quad c(\bar{x},0)=\delta(\bar{x}-\bar{x}_{0}).\label{eq:diff_eq_general}\end{equation}
Here, $D$ is the diffusion coefficient, $c(\bar{x},t)$ is the probability
density of finding the diffusing particle in an infinitesimal volume
around $\bar{x}$ at time $t$ given that it started at $\bar{x}_{0}$
at time $t=0$.

The survival probability $S(t)$ is defined as the probability that
by time $t$ the particle has not crossed the boundary of $\Omega$.
$S(t)$ can be found by integrating $c$ over $\Omega$, or by integrating
the probability flux ($D\nabla c\cdot\hat{n}$) out of $\Omega$,
\begin{equation}
S(t)=\int_{\Omega}c(\bar{x},t)d\bar{x}=1-D\int_{0}^{t}\int_{\partial\Omega}\nabla c(\bar{x},\tau)\cdot\hat{n}|dA|d\tau,\label{eq:S_t_deff}\end{equation}
where $dA$ is the element of the surface area of $\partial\Omega$.
Conversely, the exit probability per unit time (exit current) is:
\begin{equation}
p(t)=-D\int_{\partial\Omega}\nabla c(\bar{x},t)\cdot\hat{n}|dA|=-\frac{\partial S(t)}{\partial t}.\label{eq:exit_probability}\end{equation}
The above boundary and volume integral expressions are equal by the
Gauss's theorem and the diffusion equation. The probability density
for the exit location on $\partial\Omega$, i.e. the splitting probability
is: \begin{equation}
j(\bar{x},t)=D\frac{\nabla c(\bar{x},t)\cdot\hat{n}_{\bar{x}}}{-p(t)},\quad\bar{x}\in\partial\Omega.\label{eq:splitting_probability}\end{equation}
The first-passage (FP) propagation consists of sampling from the exit-time
probability $p(t)$ and the splitting probability $j(\bar{x},t)$.
The no-passage (NP) propagation entails sampling from the probability
density to find the particle near $\bar{x}$ at time $t$ under the
condition that the particle has not exited $\Omega$ by time $t$:
\begin{equation}
g(\bar{x},t)=\frac{c(\bar{x},t)}{S(t)}.\label{eq:NP_propagator}\end{equation}

\subsection{Propagators on a segment in \emph{1d}}

In $1d$, each protective domain is a line segment of length $L$.
After translation and expressing the particle position in the units
of $L$ and expressing time in the units of $L^{2}/D$, particle diffusion
on segment $[a,b]$ is described by the following equation on $[0,1]$:

\begin{equation}
\frac{\partial^{2}c}{\partial x^{2}}=\frac{\partial c}{\partial t}\label{eq:diffusion_1D_normalized}\end{equation}
with the boundary conditions $c(0,t)=c(1,t)=0$ and the initial condition
$c(x,0)=\delta(x-x_{0})$, where $x_{0}$ is the initial position
of the particle. Note that the same propagator can be used for a particle
near a boundary if the boundary is absorbing, however, for reflective
boundaries a von Neumann on one of the sides of the domain is needed.
We do not discuss such a propagator here since we exclusively use
periodic boundary conditions.

\begin{figure}
\centering{}\includegraphics[width=0.7\textwidth]{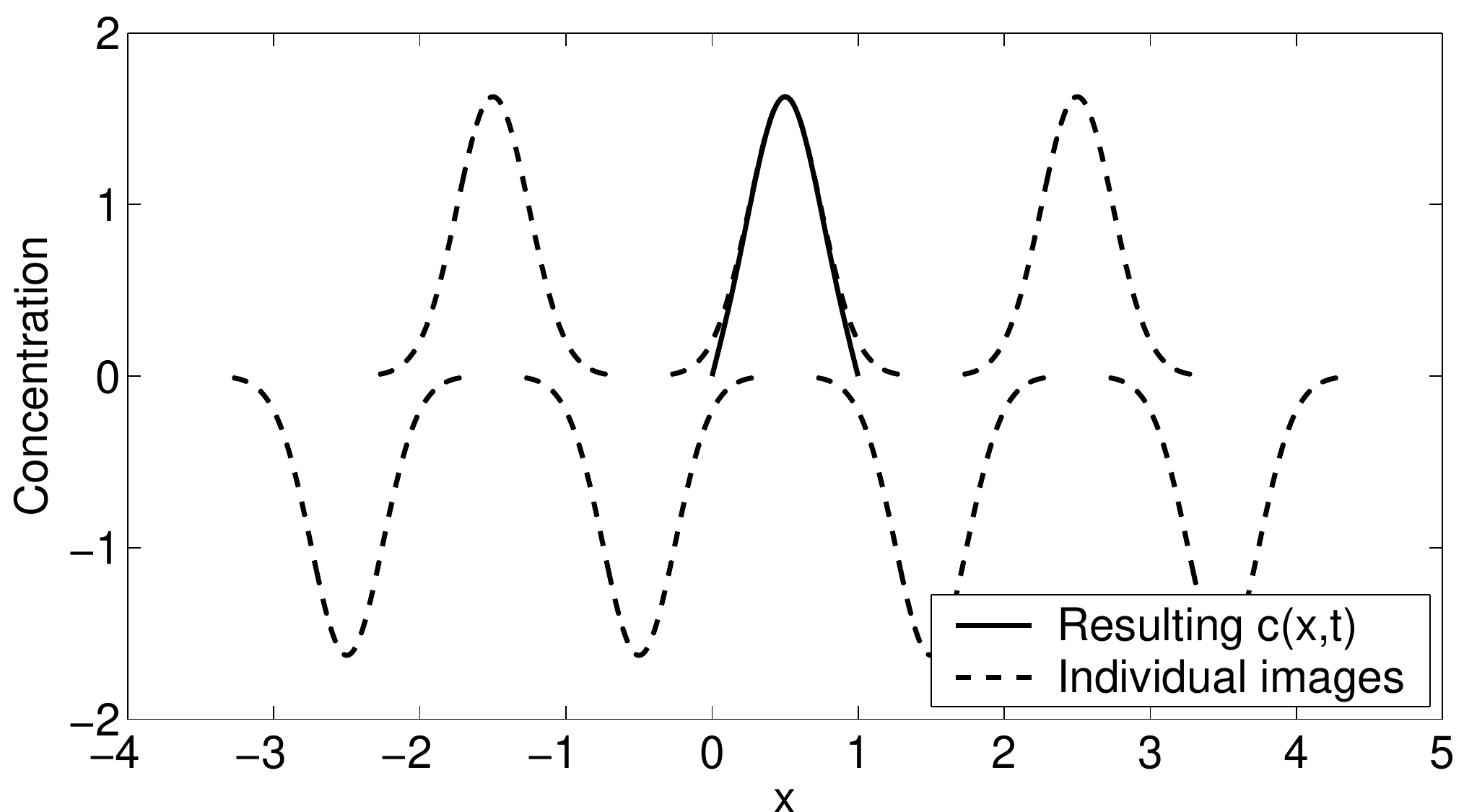}\caption{\label{fig:images}The solution for the probability density $c(x,t)$
is obtained by summing the images, signed (positive or negative) copies
of the fundamental (Gaussian) solution placed at appropriate positions
along the $x$ axis. The individual images are shown as dashed lines
and the solid line is the resulting solution. }

\end{figure}

The solution to Eq. (\ref{eq:diffusion_1D_normalized}) can be written
as the eigen-function expansion,

\begin{equation}
c(x,t)=2\sum_{k=1}^{\infty}\sin(k\pi x)\sin(k\pi x_{0})e^{-k^{2}\pi^{2}t}.\label{eq:c_x_t_long_t}\end{equation}
This series converges quickly for $t\gtrsim\frac{1}{\pi^{2}}$. An
alternative is to take advantage of the fundamental solution (Gaussian)
and express $c(\bar{x},t)$ as a sum of properly shifted imageswith
alternating positive and negative signs, as shown in Fig. \ref{fig:images},

\begin{equation}
c(x,t)=\frac{1}{\sqrt{4\pi t}}\sum_{k=-\infty}^{\infty}(-1)^{k}\exp\left\{ -\frac{\left[x-\left(k+\frac{1}{2}+(-1)^{k}(x_{0}-\frac{1}{2})\right)\right]^{2}}{4t}\right\} ,\label{eq:c_x_t_short_t}\end{equation}
which can also be derived from Eq. (\ref{eq:c_x_t_long_t}) through
the Poisson summation formula.

This expression converges quickly when $t\lesssim\frac{1}{4}$. The
needed FP and NP propagators can be now obtained by substituting either
of these two solutions into Eqs. (\ref{eq:exit_probability}), (\ref{eq:splitting_probability})
and (\ref{eq:NP_propagator}). Further technical details can be found
in Appendix B.

\subsection{\label{sub:Pair-Propagators}Pair propagators in \emph{1d}}

\begin{figure}
\begin{centering}
\includegraphics[width=0.8\textwidth]{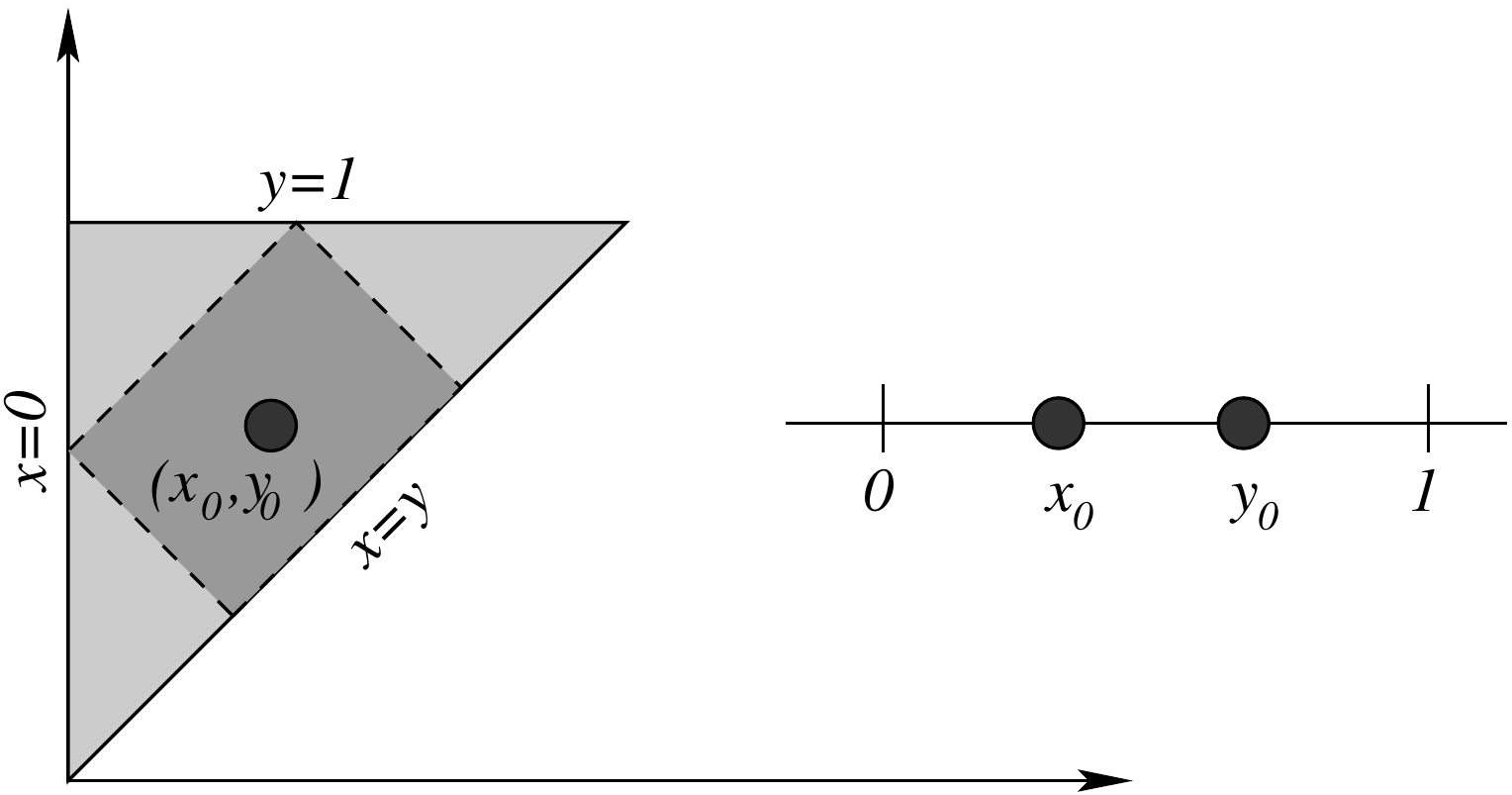}
\par\end{centering}

\caption{\label{fig:pair1D} Two particles on a line protected by a single
domain $[0,1]$. The edges on the triangle are absorbing (Dirichlet)
boundaries for the diffusion problem, and correspond to three possible
event outcomes: $x$ exits to the left, $y$ exits to the right, and
$x$ and $y$ collide.}

\end{figure}

In order to make the FPKMC algorithm more efficient and to enable
exact sampling of particle collisions, we protect close particle in
pairs. Appropriate FP and NP propagators for the protected pairs should
allow correct sampling of the collision time and particle positions.
Consider two particles at $x$ and $y$ on a unit segment $[0,1]$
so that $0<x<y<1$. To obtain the needed propagators one can solve
the following two-dimensional diffusion problem \[
\left(\frac{\partial^{2}c}{\partial x^{2}}+\frac{\partial^{2}c}{\partial y^{2}}\right)=\frac{\partial c}{\partial t},\]
with the boundary conditions $c(t,x=0,y)=c(t,x,y=1)=c(t,x,x)=0$ and
the initial condition $c(0,x,y)=\delta(x-x_{0},y-y_{0})$. This diffusion
equation has to be solved on the triangle shown in Fig. \ref{fig:pair1D}.
Absorption of the pair at the boundary $x=y$ corresponds to a collision
between the two particles.

Rather than solving this two-dimensional problem, we note that there
is a simpler diffusion problem whose solution may allow us to propagate
the pair almost as efficiently. Namely, we can use the solution of
the same equation on any domain that is entirely contained inside
the triangle. To retain the ability to sample collisions, a finite
fraction of the $x=y$ line should be included in the new domain boundary.
With this in mind, let us introduce new variables for the center of
mass $u=\frac{1}{\sqrt{2}}(x+y)$ and the difference $v=\frac{1}{\sqrt{2}}(y-x)$
and define the new domain as the maximal rectangle that can be inscribed
in the triangle so that one of its sides coincides with the collision
line $v=0$ (that is $x=y$). In these new coordinates, the two-dimensional
diffusion problem on the rectangle separates into two \emph{1d} problems,
one for $u$ and one for $v$, with the absorbing boundary conditions
on all rectangle sides. This is convenient since one can use the same
FP and NP propagators already derived for a unit segment in \emph{1d}.
First, one samples two exit times, one for {}``walker'' $u$ and
another for {}``walker'' $v$. The exit time out of the inscribed
rectangle is the shorter of the two. The exit coordinates are sampled
using the splitting FP probability \emph{j} for the {}``walker''
whose exit time is shorter and using the NP propagator for the other
{}``walker''. The two particles collide when {}``walker'' $v$
exits to $v=0$. All other outcomes correspond to pair propagation.
Note that this algorithm preserves exact statistics of diffusive propagation
and collisions of the protected pair. The sampled time increments
are somewhat smaller on average than could be achieved by using the
full triangular domain, which is an acceptable cost to pay for eliminating
the need to compute and sample from the more complicated\emph{ 2d}
propagators on the triangular domain.

\subsection{\label{sub:higher_d}Generalization to higher dimensions}

For the case of isotropic diffusion in any number of dimensions $m$,
the use of hyper-rectangles or hyper-cubes for protecting the particles
is convenient because the diffusion equation separates into \emph{m}
one-dimensional diffusion equations, one for each Cartesian direction.
The same holds for anisotropic diffusion provided the edges of the
protective hyper-rectangles are oriented along the principal axes
of the diffusion tensor. In both cases the FP and NP probability distributions
for \emph{m} dimensions are the products of \emph{m} one-dimensional
distributions. Therefore, one can use the one-dimensional propagators
to sample time and location of exit out of the protective hyper-rectangle.
To do this, \emph{m} exit times are sampled from the corresponding
\emph{m} one-dimensional FP propagators and the shortest of them,
say $t_{k}$, is taken as a sample of the exit time. Then the splitting
probability function $j(t_{p})$ is used to sample the exit location
for the Cartesian direction $k$ and the NP distributions at $t_{k}$
are sampled to obtain the walker exit position for the remaining $m-1$
Cartesian directions.

A similar method can be used to propagate protected pairs and possibly
larger groups of particles. For example, for the case of a pair of
square-shaped particles protected by a square in $2d$, the change
of variables is used to transform the problem to diffusion on two
rectangles, one for each Cartesian coordinate. Accordingly, four one-dimensional
FP propagators are used to sample an exit time $t_{p}$ and an exit
dimension \emph{k} in the transformed coordinates. Then, three NP
propagators are used to sample particle positions at the exit time
$t_{p}$ in the three other dimensions. Using the splitting probability
function $j(t_{p})$ one decides if the sampled exit indicates a collision
along one of the two Cartesian directions. If so, whether or not the
pair has actually collided is determined by the particle separation
along the other Cartesian direction. For a collision to occur, the
latter should be smaller than the sum of the half-widths of the square-shaped
particles.

\subsection{Sampling}

Given a random number $r$ uniformly distributed on $r\in[0,1)$,
a sample first passage time $t_{p}$ is obtained by solving $S(t_{p})=r$
or simply as $t_{p}=S^{-1}(r)$, where $S(t)$ is the survival probability
function on domain $\Omega$. An exit location sample $x\in\delta\Omega$
can be obtained from the splitting probability density $j(\boldsymbol{x},t_{p})$.
Splitting probabilities for exit to the ends of a segment in \emph{1d}
are given by two numbers $j_{1}(t_{p})$ and $j_{2}(t_{p})$, $j_{1}(t_{p})+j_{2}(t_{p})=1$.
When the initial position of the walker is at the center of a protective
segment, the splitting probabilities are equal and independent of
the first passage time, $j_{1}=j_{2}=\frac{1}{2}$. Thus, using protective
segments (or hyper-rectangles) concentric with the initial particle
positions is particularly convenient. Given a NP probability density
function $g(x,t)$, a sample of the no-passage position $x_{np}$
at time $t'$ can be obtained by solving $G(x_{np};t')=r$, or $G^{-1}(r;t')=x_{np}$,
where \emph{r} is a random number uniformly distributed on $r\in[0,1)$,
$G(x_{np};t')=\int_{_{x_{1}}}^{x_{np}}g(x,t')dx$ is the cumulative
NP distribution function at time $t'$.

Even for the simple case of diffusion on a \emph{1d} segment with
two absorbing ends, no closed form solution exist and the FP and NP
propagators are available only in the form of series expansions. We
have implemented and tested two techniques for sampling FP and NP
propagators, both taking advantage of the fast convergence of the
expansion series. The first technique uses pre-tabulated propagators
or, rather, appropriate inverse functions for fast lookup and interpolation.
This method is particularly simple for the case of continuous diffusion
on a segment in \emph{1d} because the propagators can be stored as
one-dimensional (FP) or two-dimensional (NP) tables (the time and
position variables in the propagator tables are stored in the units
of $L^{2}/D$ and $L$, respectively).

The second technique is rejection sampling that relies on a converging
series of upper and lower bounds to exactly sample from the FP and
NP distribution density functions at a the cost of an occasional rejection.
The needed series of bounds is obtained by integrating the series
expressions (\ref{eq:c_x_t_long_t}) and (\ref{eq:c_x_t_short_t})
and observing that the terms in the resulting series solutions for
the propagators have alternating signs and absolute values that monotonically
decrease with the increasing term order. Thus, subsequent partial
sums of the alternating series present an alternating sequence of
increasingly tight upper and lower bounds. Taking advantage of particularly
fast convergence of series (\ref{eq:c_x_t_short_t}) and (\ref{eq:c_x_t_long_t})
for short ($t\lesssim\frac{1}{4}$) and long ($t\gtrsim\frac{1}{\pi^{2}}$)
times, respectively, it is possible to construct very tight bounds
to the exact distribution functions. As a result, rejection sampling
is efficient because rejections are infrequent and it rarely takes
more than two bound evaluations to accept or reject a sample. This
sampling technique is especially useful when it is difficult or impossible
to evaluate and/or invert the cumulative probability distribution
functions. However, when the inverse distributions are available,
such as in the form of look-up tables, we found both table lookup
and rejection techniques similarly efficient.

Further technical details on computing and sampling the $1d$ propagators
are given in Appendices \ref{AppendixA} and \ref{sec:Continuous-propagators}.

\section{Computational tests on accuracy and efficiency}

In this section we apply the FPKMC algorithm to several model diffusion-reaction
problems as a way to validate the method and compare its efficiency
to traditional algorithms. The first test is for one species annihilation
in both $1d$ and $3d$, and the second test is for two-species annihilation
in $3d$.

\subsection{Annihilation in $1d$}

As a first test of our algorithm we study the kinetics of a diffusion-controlled
reaction of particle annihilation $A+A\Rightarrow0$ in $1d$. The
simulation starts with a large number of particles in a periodic box
and proceeds with a steady decline in the number of particles as they
annihilate. The FPKMC algorithm is ideally suited for this type of
problems because it adaptively adjusts the effective time step and
hop size during the course of the simulation. In the beginning of
the simulation the protective regions and the time steps between successive
events are small. As the simulation proceeds the mean free path increases
and larger hops are taken. This allows one to simulate the process
all the way to complete annihilation of all particles without expending
significantly more computational effort per particle at lower densities,
in stark contrast to the traditional algorithms.

Exactly how the new method's efficiency compares to traditional hop-by-hop
KMC with varying particle density depends on dimensionality and, possibly,
on whether or not the walks are continuous or discrete. For the case
of continuous diffusion in $1d$, simulation efficiency is manifestly
independent of particle separation. In higher dimensions, efficiency
of the FPKMC method diminishes with decreasing density. Luckily, the
need for efficient and accurate simulations of diffusion-reaction
processes is limited to dimensions $d<4$ because for $d\geqslant4$
the effect of correlations can be neglected and the mean field kinetics
becomes accurate \citet{A_plus_B_annihilation}.

Efficiency of Monte Carlo simulations of diffusion-controlled annihilation
or coalescence reactions can be further enhanced by periodic replication.
Although not directly related to the methods discussed in this paper,
periodic replication is especially efficient when used in combination
with the FPKMC method. In has been noted in the past \citet{A_plus_B_annihilation2}
that in the reactive systems where the number of particles steadily
decreases, the simulated kinetics of the late stages of the reaction
are not representative of the kinetics of interest. This is due to
an inevitable growth of correlations among particles remaining in
the box as the simulation progresses. In annihilation and coalescence
reactions, correlations develop because the surviving particles are
less likely to have reacting neighbors in close proximity. This {}``correlation
hole'' effect spaces the particles more evenly than in a random configuration,
thus affecting the distribution of times of particle collisions. If
and when the growing correlation length becomes comparable to the
simulation box size, the kinetics becomes distorted, as a finite size
effect. Reduction in the number of surviving particles compounds the
difficulties making the tail kinetics noisy. For these reasons, the
last 2-3 decades of simulated kinetics are typically discarded \citet{A_plus_B_annihilation2}.
On the other hand, the first few decades of the simulated kinetics
are often discarded because they reflect more the initial particle
distribution (usually random) rather than the reaction kinetics of
interest. Combined with the usual limit on the number of particles
(typically $10^{6}$), these unwanted behaviors limit to 4-6 decades
the useful time interval over which one can observe and quantify diffusion-reaction
kinetics.

\begin{figure}
\begin{centering}
\includegraphics[width=0.8\textwidth]{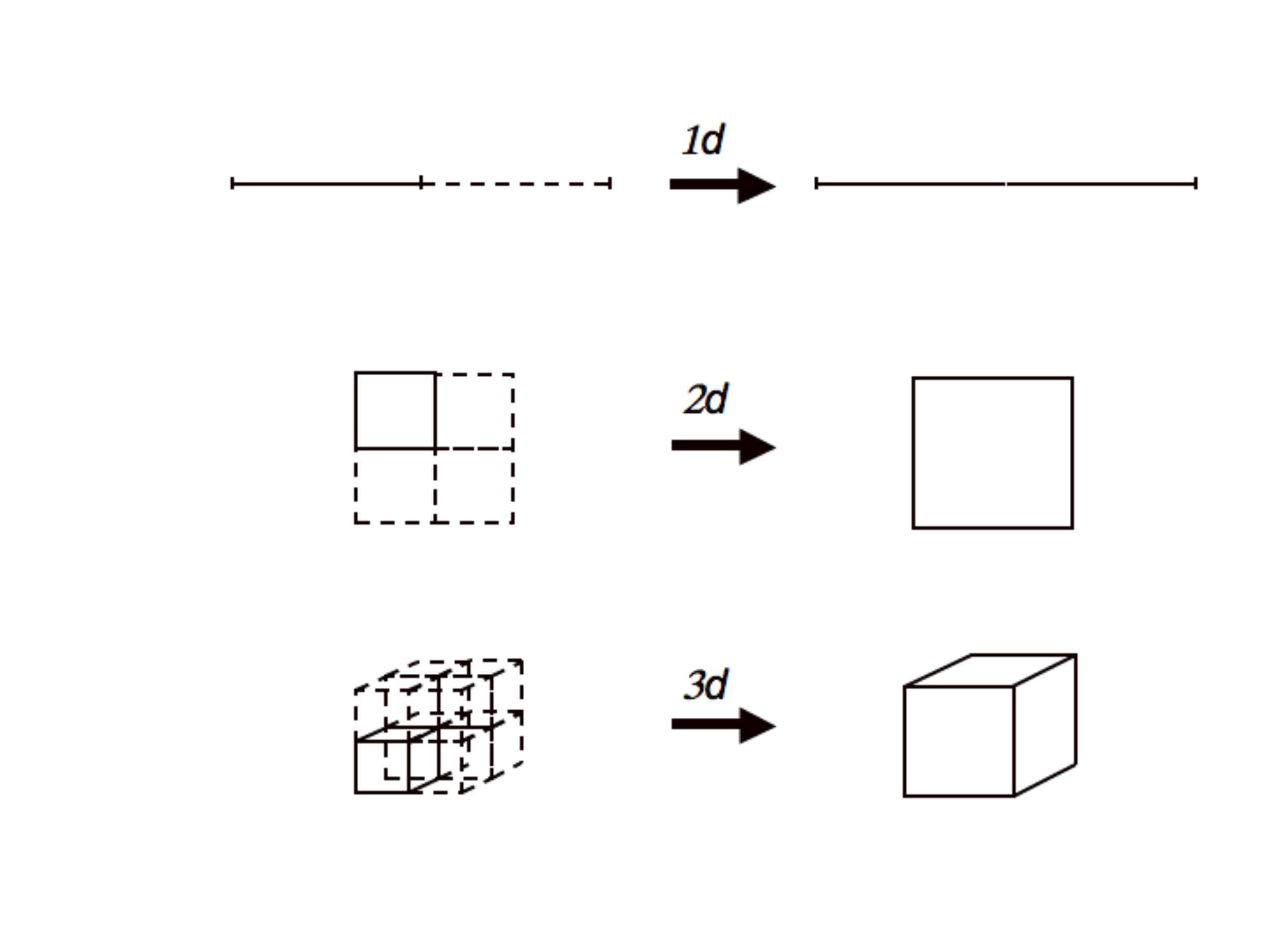}
\par\end{centering}

\caption{\label{fig:replication}Schematic of the periodic replication procedure. }

\end{figure}

Periodic replication works as follows \citet{PeriodicReplication_KMC}.
First, the simulation starts from its initial configuration in a periodic
box and proceeds as usual. Then, once the correlation length grows
comparable to the box size, the particles are synchronized to the
current point in time and $2^{d}$ neighboring periodic replicas are
combined in a new box double the linear size of the old box, see Fig.
\ref{fig:replication}. The simulation restarts in the new box in
which the particles that were previously periodic image slaves of
each other are now treated independently. Such box doubling should
be repeated whenever the correlation length approaches a fraction
of the current box size. Given that correlations propagate by diffusion,
we estimate that the correlation length should grow as $\sim\sqrt{Dt}$
where $D$ is the diffusion coefficient. Thus, the volume should be
replicated whenever $\sqrt{Dt}$ grows to become an appreciable fraction
of the linear box size $L$. Hence, the physical time $t$ elapsed
between replications should at least quadruple with each replication.
Assuming that in the interval between two box replications the number
of surviving particles decreases as $t^{-\alpha}$, each replication
increases the number of particles by a factor $2^{d-2\alpha}$.

\begin{figure}
\begin{centering}
\includegraphics[width=0.75\textwidth]{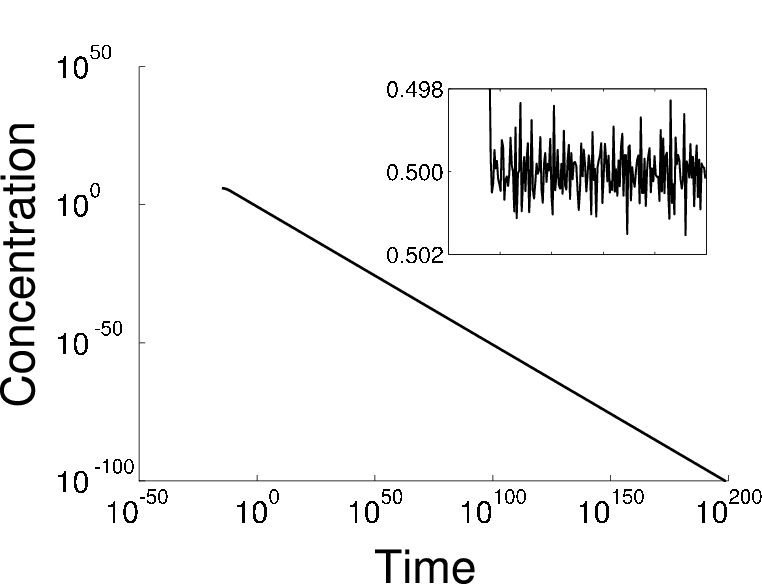}
\par\end{centering}

\caption{\label{fig:superlong}Kinetics of $A+A\rightarrow0$ annihilation
reaction in $1d$, using replication. The inset is the logarithmic
derivative of the kinetic curve taken over the same time interval
reproducing the theoretical exponent $\alpha=0.5$ for this reaction.}

\end{figure}

Remarkably, for $A+A\Rightarrow0$ and $A+A\Rightarrow A$ reactions
in \emph{1d}, $\alpha=\frac{1}{2}$ \citet{DiffusionLimitedAnnihilation}
and the doubling in the number of particles caused by each replication
is compensated by the reduction by half caused by annihilation or
coalescence taking place between two replications. Thus, in these
particular cases replications do not cause the number of particles
to grow and can continue indefinitely. Combined with the fact that
FPKMC efficiency does not depend on particles density in $1d$, we
can simulate such processes to an arbitrarily long physical time,
as shown in Fig. \ref{fig:superlong}.

In other cases when $d-2\alpha>0$, periodic replications will result
in increases in the number of particles. For example, for the case
$d=2,$ $\alpha=\frac{3}{4}$, the number of particles is expected
to double after two replications, whereas for the case $d=3,$ $\alpha=1$,
the number of particles will double after each replication. Therefore,
sooner or later the number of particles will grow too large to continue.
Thus, rather than follow the common practice to start from a maximum
size that fits into memory and then simulate the reaction to the end,
it may be better to start from a small number of particles and let
the system grow by replications to a maximum size afforded by the
computer memory. Assuming that with FPKMC we can handle simulations
with at most $10^{9}$ particles and that is is safe to start with
just $10^{3}$ particles in the box, this allows as many as $\log_{2}(\frac{10^{9}}{10^{3}})\approx20$
replications for the $d=3,$ $\alpha=1$ case and $40$ replications
for the $d=2,$ $\alpha=\frac{3}{4}$ case. Thus, replications should
allow extension of the useful time horizon of such reaction-diffusion
simulations to $4^{20}$ (12 decades of time) and $4^{40}$ (24 decades
of time), respectively. Obviously, this recipe also eliminates the
earlier mentioned tail effects since correlations are never allowed
to catch up with the growing box size and the number of particles
in the end of the simulation is large%
\footnote{Even though the plots presented in Figures 5--8 appear rather smooth,
FPKMC simulations faithfully account for and contain wealth of statistical
information on fluctuations and correlations in the considered diffusion-reaction
models. If desired, such statistics can be extracted from the same
simulations. %
}.

\subsection{Annihilation in $3d$}

\begin{figure}
\begin{centering}
\includegraphics[width=0.5\textwidth]{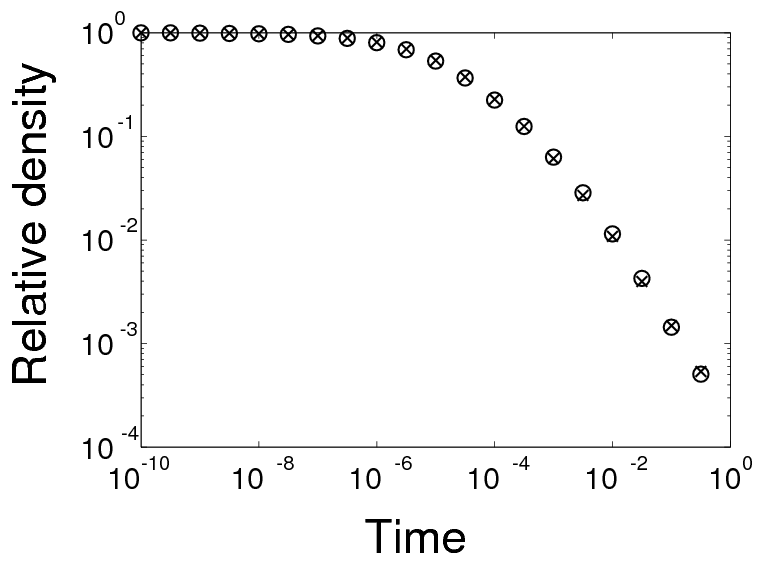}
\par\end{centering}

\caption{\label{fig:3d_comparison}Comparison between the standard KMC calculations
(crosses) and the first-passage KMC (circles) results for $A+A\rightarrow0$
annihilation reaction in $3d$. Each of the two curves is the average
over one thousand simulations each starting with $8000$ particles.}

\end{figure}

The next test is a simulation of a diffusion-controlled reaction of
particle annihilation $A+A\Rightarrow0$ in $3d$. Figure \ref{fig:3d_comparison}
compares the annihilation kinetics simulated using the FPKMC method
and a standard KMC algorithm in which the particles are propagated
by small hops \citet{BIGMAC,LAKIMOCA_OKMC,Smoldyn_v1}. Each simulation
starts with 8000 cube-shaped particles occupying initially a volume
fraction of 0.004 of the simulation cube volume with periodic boundary
conditions applied in all 3 dimensions. The two kinetics are seen
to be identical within small statistical errors.

\begin{figure}
\begin{centering}
\includegraphics[width=0.8\textwidth]{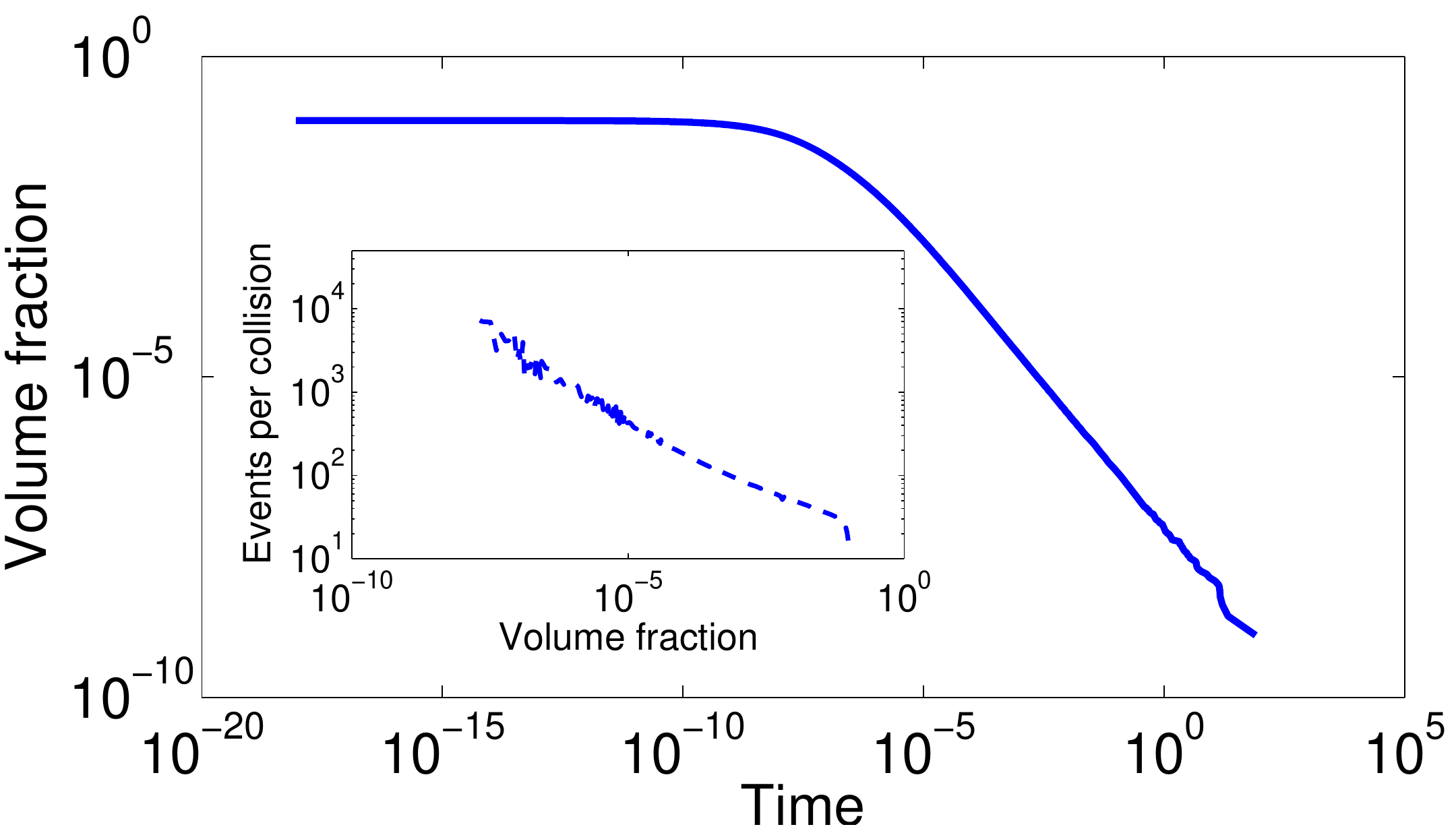}
\par\end{centering}

\caption{\label{fig:125mil}Simulated kinetics of $A+A\rightarrow0$ reactions
in $3d$. This curve is for a single run starting with $216\cdot10^{6}$
particles at a volume fraction of $0.1$ (the reaction completes in
79 CPU hours). The plot in the inset shows that even in an FPKMC simulation
the number of events (hops) per collision can become large at very
low particle densities.}

\end{figure}

The greater efficiency of the FPKMC method allows simulations of very
large numbers of diffusing and reacting particles at a modest computational
effort. An example is shown in Fig. \ref{fig:125mil} for the same
annihilation reaction in $3d$ but starting with 216 million particles.
The reaction completes in a few CPU days on a modest workstation.
By comparison, we estimate that it would take tens of CPU years on
the same workstation to complete this simulation using the standard
KMC algorithm %
\footnote{The code we refer to here as {}``standard KMC algorithm'' is $BigMac$
\citet{BIGMAC} that has been extensively used for simulations of
various diffusion-reaction processes. \emph{BigMac} has not been specifically
optimized for conditions of low particle density. %
}. 

Each step of the FPKMC algorithm requires more calculations than one
step of the standard KMC algorithm. The obvious overhead due to the
need to sample from the more complicated distributions is relatively
minor whereas more serious amount of computational effort in FPKMC
is spent on keeping track of near neighbor particles, space partitioning
and other tasks associated with particle protection, as well as on
maintaining the event queue after every event. Note that all of these
elements appear in other KMC algorithms in one form or another, and
therefore standard techniques can be used. FPKMC codes used for simulations
presented in this paper have not been extensively optimized although
some of the more obvious inefficiencies have been addressed. Nevertheless,
it should be of interest to compare the net efficiency of the FPKMC
simulations to that of the standard KMC method in the units of CPU
time per particle collision. 

\begin{figure}
\begin{centering}
\includegraphics[width=0.5\textwidth]{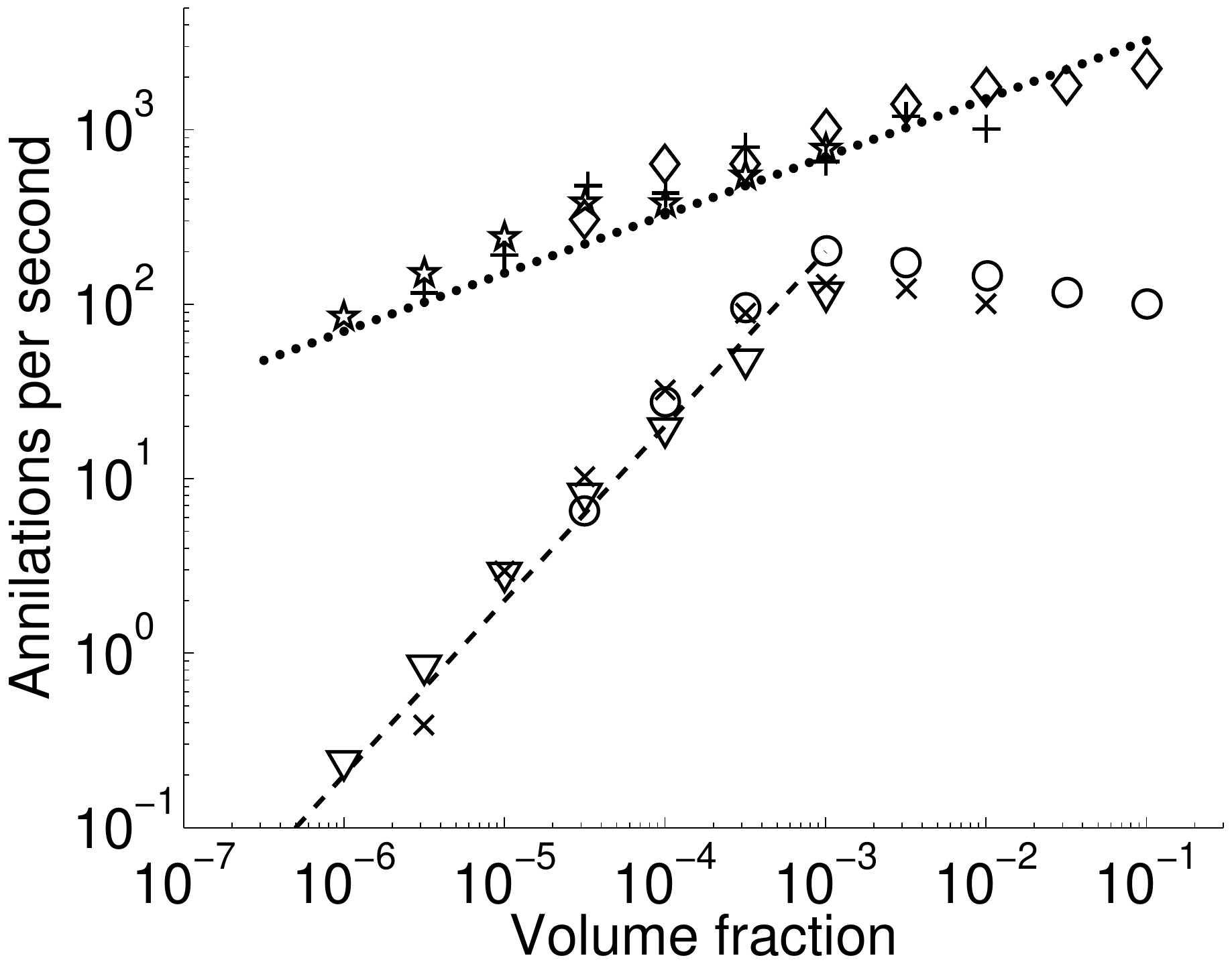}
\par\end{centering}

\caption{\label{fig:desn-scaling}Computational performance as a function of
particle density measured in six independent realizations of $A+A\rightarrow0$
annihilation reaction in \emph{3d} performed on a single CPU workstation
using the standard KMC calculations with a finite hop distance (lower
symbols) and the FPKMC algorithm (upper symbols). The dashed and dotted
lines are fitted lines with slope $\approx1$ for the standard KMC
calculations and slope $\approx\frac{1}{3}$ for the new algorithm. }

\end{figure}

A relevant comparison is given in Fig. \ref{fig:desn-scaling}, where
the number of particle collisions (annihilation) per second of CPU
time is plotted as a function of particle density for two series of
simulations of $A+A\Rightarrow0$ annihilation reaction in \emph{$3d$}
using the FPKMC and the standard KMC algorithms. Each of the two series
consists of three simulations starting from the same high initial
volume fraction of particles of $0.05$ and ending at a much lower
volume density $10^{-6}$ (the simulations proceed from right to left).
Because in the beginning some of the particles are very close to each
other, it is necessary to use very small hop sizes in order to ensure
that collision sequences are properly resolved. In the course of the
simulation the nearest neighbor particle pairs progressively annihilate
and the time step gradually increases. Eventually, as the average
particle spacing gradually increases due to continued annihilation,
the average number of hops between any two collisions also increases
and efficiency of the standard method deteriorates inversely proportionally
to the particle volume fraction (density) $\rho$. At the same time,
FPKMC automatically selects the propagation step size to meet the
local geometrical requirements, achieving an exact solution without
any tuning. Efficiency of the FPKMC algorithm is proportional to $\rho^{-\frac{1}{3}}$
throughout the whole range of simulated particle densities. Thus,
while FPKMC is competitive with the standard method even at high particle
densities, at a sufficiently low density the new algorithm is certain
to outperform the standard method. 

\subsection{Two-species annihilation in $3d$}

The last computational test we report in this paper is a simulation
of $A+B\Rightarrow0$ annihilation reaction in \emph{$3d$.} In this
reaction, particles do not interact with particles of its own kind
but annihilate on collisions with unlike particles. As has been first
observed in Refs. \citet{A_plus_B_annihilation,A_plus_B_annihilation_theory},
when the numbers of A and B particles are close to the stoichiometry
(50:50), this reaction does not follow the mean field asymptotic kinetics
$t^{-1}$ but rather $t^{-\frac{3}{4}}$ (most other diffusion-reaction
systems follow the mean field behavior for $d>2$). This peculiar
scaling was attributed to the emergence and growth of alternating
A-rich and B-rich domains that effectively limit the annihilation
reactions to inter-domain boundaries.

\begin{figure}
\begin{centering}
\includegraphics[width=0.75\textwidth]{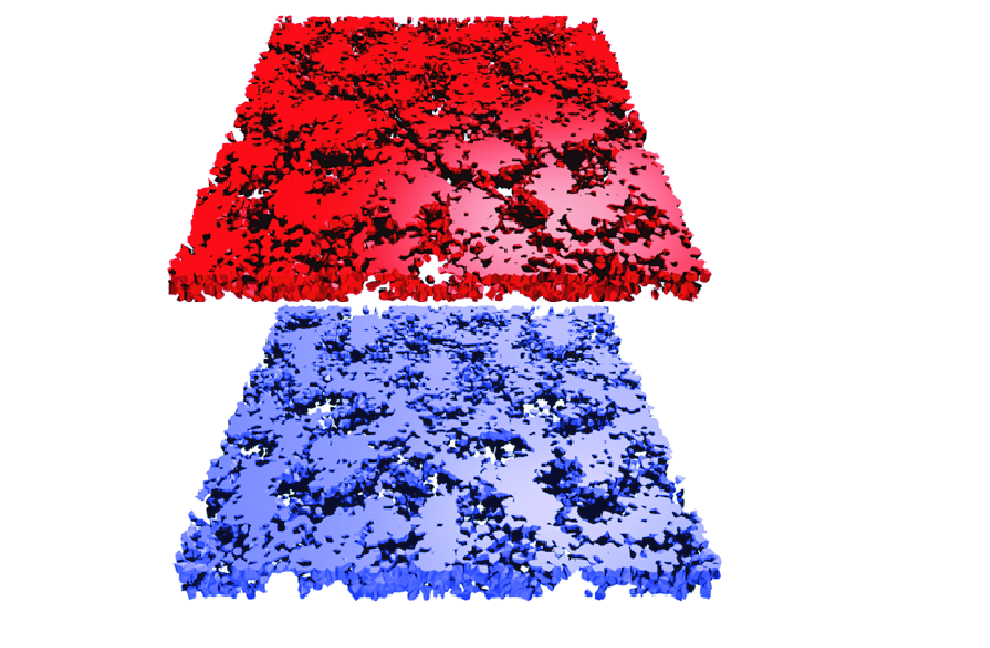}
\par\end{centering}

\caption{\label{fig:AB-domains}A thin slice through the domain structure formed
in a simulation of $A+B\rightarrow0$ reaction in \emph{3d}. The boundaries
of the A-rich (red) and B-rich (blue) domains are identified with
the sides of Voronoi polyhedra shared by unlike particles. The two
domains are complementary and fill the space when brought together.}

\end{figure}

An important physical realization of such a situation is recombination
of vacancies and interstitials produced in crystal by neutron, ion
or electron irradiation \citet{LAKIMOCA_OKMC}. Here we limit our
study to a model system in which particles A and B are cubes with
the same size and diffusion coefficients and react only with particles
of the opposite species. The slow-down caused by domain growth combined
with steadily decreasing particle density makes standard KMC simulations
particularly inefficient which has so far prevented quantitative investigations
of reaction kinetics and domain geometry in such systems, especially
in $3d$. The FPKMC method handles this reaction with relative ease
in arbitrary dimensions. Figure \ref{fig:AB-domains} shows the geometry
of a thin slice through the domain configuration produced in a FPKMC
simulation of $A+B\Rightarrow0$ reaction in $3d$ starting with $10^{6}$
particles (no replication). 

\section{Summary}

We have developed the method of First-Passage Kinetic Monte Carlo
(FPKMC) for simulations of diffusion-reaction processes. By partitioning
the space into non-overlapping protective domains around each particle
and/or particle pair, the $N$-body problem of collisions among $N$
Brownian particles or random walkers is factorized into $N$ single-body
problems or, alternatively, \emph{$K_{1}$} single body and $K_{2}$
two-body problems, $K_{1}+2K_{2}=N$. Rather than performing small
diffusional hops, exact solutions for first-passage and no-passage
statistics are used to propagate the particles inside the domains.
On each Monte Carlo cycle a single particle or a single particle pair
is propagated to the boundary of its protective domain. This is sometimes
followed by a no-passage propagation of one or few neighboring particles
or pairs.

The resulting algorithm is event-driven and asynchronous: each protected
particle or pair propagates in its own spatial domain, from its own
spatial and time origin and following its own propagation time clock.
The new method remains efficient at low densities because only one
or a few particles are propagated on every cycle over distances close
to the inter-particle spacing. The FPKMC method is exact for a wide
class of diffusion-reaction models in which Brownian particles or
random walkers do not interact until they collide (hard-core models).
The accuracy and efficiency of the new method is demonstrated in simulations
of several well-studied diffusion-reaction models that have previously
presented serious computational challenges for Monte Carlo simulations.

We would like to emphasize that the FPKMC method focuses on bringing
the particles close to collisions leaving aside the nature of reactions
taking place on collisions. Thus, although statistics of first passage
processes finds its uses in efficient handling of the reaction events
\citet{ReversiblePairReaction}, such issues are outside the scope
of this paper in which we consider only the simplest collision outcomes
- annihilation and coalescence. Extension of the FPKMC method to simulations
of more complicated reaction kinetics in which diffusional propagation
takes place simultaneously with other competing stochastic processes
will be presented elsewhere.

The asynchronous and adaptive nature of the FPKMC algorithm enables
the method to effectively deal with stiff diffusion-reaction problems
in which rate processes with vastly differing time scales coexist
and compete. Examples of such situations include the occurence of
fast diffusion of adatoms versus slow diffusion of adatom clusters
in crystal growth simulations and fast diffusion of interstitials
versus slow diffusion of vacancies in radiation damage in metals.
An extension of the FPKMC method to stiff reaction-diffusion systems
will be presented in a forthcoming publication.

\section{Acknowledgments}

This work was performed under the auspices of the U.S. Department
of Energy by Lawrence Livermore National Laboratory under Contract
DE-AC52-07NA27344. This work was supported by the Office of Laboratory
Directed Research at LLNL and the Office of Basic Energy Sciences
U. S. Department of Energy. The authors would like to express their
gratitude to G. Martin, S. Redner, W. Cai, C. Mailhiot, F. Willaime,
M.-C. Marinica and T. Diaz de la Rubia for fruitful discussions.

\section{Appendices }

\appendix

\section{\label{AppendixA}Sampling from series expansions}

The distributions sampled in FPKMC are given in the form of series
expansions. This appendix describes a general rejection technique
and its application to sampling from such series expansions. In the
following appendix \ref{sec:Continuous-propagators}, first-passage
(FP) and no-passage (NP) propagators suitable for this technique are
derived for the case of continuous diffusion. FP and NP propagators
for discrete random walks on lattices will be presented in forthcoming
publications.

Here by a distribution $c(x)$ we mean a function that is non-negative
everywhere and whose integral is bounded, i.e. $c(x)\ge0$ and $\int c(x)dx<\infty$.
Sampling from $c(x)$ means drawing random numbers $x$ distributed
according to the probability density $c(x)/\int c(x')dx'$. Rejection
sampling is often used when it is difficult to invert the cumulative
distribution, e.g. to solve $\xi=\intop^{x}c(x')dx'$ for $x$, and,
at the same time, a majoring distribution $C(x)$ exists such that
it is easy to sample and $C(x)\ge c(x)$ for all $x$. Rejection sampling
proceeds as follows:
\begin{enumerate}
\item Let $x_{trial}$ be a sample of $C$. 
\item Pick a uniformly distributed random number $0\le y<C(x_{trial})$.
\item If $y<c(x_{trial})$, accept $x_{trial}$ as a sample of $c$, otherwise
reject $x_{trial}$ and go to step 1.
\end{enumerate}
If $\int_{0}^{\infty}C(t)dt/\int_{0}^{\infty}c(t)dt-1$ is small,
rejection is infrequent and the resulting sampling is efficient while
still exact in the sense of sampling from the true distribution $c(x,t)$.

When the distribution $c$ is available in the form of a converging
series expansion, the partial sums of the series can be used in the
acceptance/rejection test in step 2 of the rejection sampling algorithm
above. Suppose $c(x)=\sum_{k=0}^{\infty}c_{k}(x)$, define the partial
sums $S_{m}=\sum_{k=0}^{m}c_{k}(x)$ and assume that upper and lower
bounds $U_{m}$ and $L_{m}$ of the remainder term are available,
so that $L_{m}\le c(x)-S_{m}\leq U_{m}$ and both bounds become tighter
with each added term. Then, in step 3 above, if $y<S_{m}+L_{m}$ the
sample is accepted without evaluating any terms beyond $m$. Conversely,
if $y\geq S_{m}+U_{m}$ the sample is rejected. If however $S_{m}+L_{m}<y\leq S_{m}+U_{m}$
no decision can be made and the next order terms have to be calculated
in order to repeat the test with the same sample $y$ but using $S_{m+1}$,
$U_{m+1}$ and $L_{m+1}$. Especially simple is the case when the
sign of the remainder term $c(x)-S_{m}$ is known; the sample is accepted
if $y<S_{m}$ and the remainder is positive and rejected if $y\ge S_{m}$
and the remainder is negative. Our experience with the FPKMC algorithm
suggests that the rejection procedure requires computing only two
terms on average before the sample is accepted or rejected.

\section{\label{sec:Continuous-propagators}Propagators for continuous diffusion}

Here we derive the first passage and no passage propagators suitable
for rejection sampling on one dimensional line segments for the case
of continuous diffusion.

\subsection{First-passage propagator}

\begin{figure}
\centering{}\includegraphics[width=0.7\textwidth]{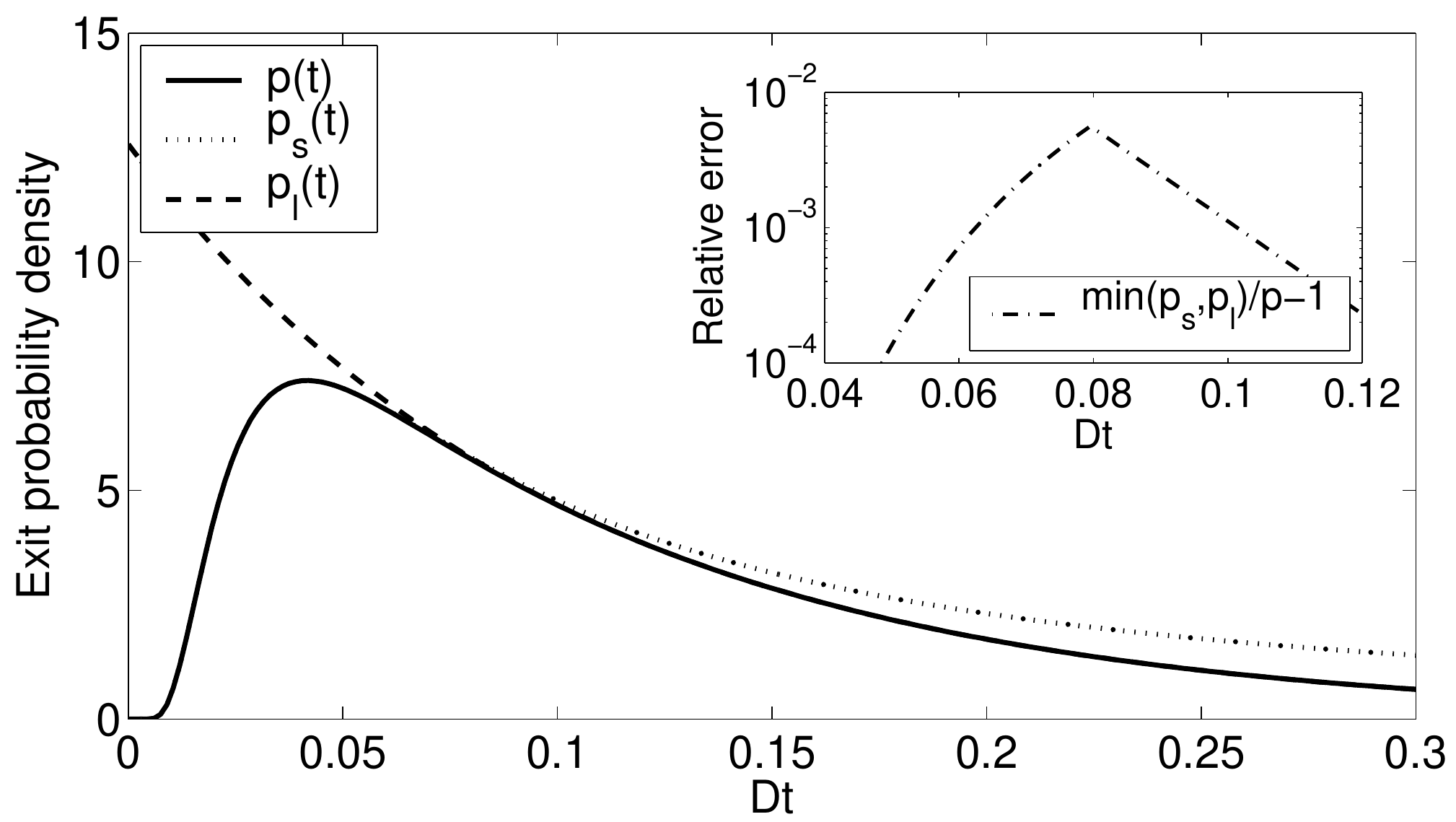}\caption{\label{fig:Comparison-of-one-term}Comparison of one-term expansions
versus full probability density. The inset shows the relative error
of using only the proposed majoring function without subsequent rejection
sampling.}

\end{figure}

Assume $c(x,t)$ is the solution to the diffusion equation given in
section \ref{sec:The-propagators}. From Eq. (\ref{eq:exit_probability})
it follows that the exit probability per unit time is $p(t)=\frac{\partial c}{\partial x}(0,t)-\frac{\partial c}{\partial x}(1,t)$.
Taking the solution in the form of the image series suitable for short
times {[}c.f. Eq. (\ref{eq:c_x_t_short_t}){]} we obtain: \[
p(t)=\frac{2\pi}{\sqrt{4\pi t}^{3}}\sum_{j=-\infty}^{\infty}(-1)^{j}\left\{ \left(j+\frac{1}{2}\right)e^{-\left(j+\frac{1}{2}\right)^{2}/4t}-\left(j-\frac{1}{2}\right)e^{-\left(j-\frac{1}{2}\right)^{2}/4t}\right\} .\]
 Retaining the two most significant terms in this series yields an
approximate expression accurate for short times \[
p_{s}(t)=\frac{4\pi}{\sqrt{4\pi t}^{3}}e^{-\frac{1}{16t}}.\]
Similarly, retaining the first term in the long time expansion (\ref{eq:c_x_t_long_t})
\[
p(t)=\sum_{k=1}^{\infty}2k\pi\sin(k\pi x_{0})[1-(-1)^{k}]e^{-k^{2}\pi^{2}t},\]
and using $x_{0}=1/2$, we obtain an approximation that is accurate
for long times, \[
p_{l}(t)=4\pi e^{-\pi^{2}t}.\]
In turns out that $p_{s}>p$ and $p_{l}>p$ for all times $t$ and,
furthermore, $p_{s}$ and $p_{l}$ intersect at $\tau_{0}\approx0.0796$.
Thus, $\min(p_{s},p_{l})$ is a tight majoring function for the true
distribution $p$ that it is accurate to within $0.6\%$ for all $t$,
as shown in Fig. \ref{fig:Comparison-of-one-term}.%

Defining the integrals

\[
\begin{array}{l}
F_{l}(\tau)=\int_{\tau}^{\infty}p_{l}(t)dt=\frac{4}{\pi}e^{-\pi^{2}\tau}\\
F_{s}(\tau)=\int_{0}^{\tau}p_{s}(t)dt=2\left[1-\textrm{erf}\left(\frac{1}{\sqrt{16t}}\right)\right]\end{array},\]
a sample of the exit time $t_{trial}$ from the majoring distribution
is \[
t_{trial}=F_{s}^{-1}(r)=\frac{1}{16\left[\textrm{erf}^{-1}\left(1-\frac{r}{2}\right)\right]^{2}},\]
if $r<F_{s}(\tau_{0})$, and \[
t_{\textrm{trial}}=F_{l}^{-1}(r-F_{s}(\tau_{0}))=\tau_{0}-\pi^{-2}\log\frac{r-F_{s}(\tau_{0})}{F_{l}(\tau_{0})},\]
otherwise. Here, $r$ is a random number uniformly distributed on
the segment $0<r\leq F_{s}(\tau_{0})+F_{l}(\tau_{0})$. Using this
trial value of the exit time, it is now straightforward to employ
the rejection sampling technique described in appendix A. For the
short time series, the terms of the series alternate in sign and decrease
in magnitude with increasing \emph{$m$} so that the partial sums
\emph{$S_{m}$} provide an alternating sequence of upper and lower
bounds. The same holds for the long time series, provided $\tau_{trial}\ge\frac{1}{18\pi^{2}}\approx0.0056$.
Depending on the value of $t_{trial}$, one or the other series converges
faster. It is possible to choose among the two alternatives by comparing
$t_{trial}$ to $\tau_{0}$. However, it is more efficient to use
a different switchover time $t_{switch}$ that optimizes the computational
cost of the sampling routine. For our implementation $t_{switch}\approx0.033$
turned out to be optimal.

\subsection{No-passage propagator}

The no-passage propagator is the distribution density of particle
positions at time $t$ conditioned on the fact that the particle has
not reached the boundary of its protective domain by that time. As
is the case for the FP propagator described above, sampling from the
NP distribution is most efficiently done using two different expansion
series at short times and at long times. Using the sampling procedure
described below, rejections are infrequent (less than 1\%) requiring
on average less than two terms in the series expansions to accept
or reject the sample. 

For short times, the probability density $c(x,t)$ is best represented
by the image sum {[}c.f. Eq. (\ref{eq:c_x_t_short_t}){]}. The $m=0$
term of this expansion is the fundamental solution $C(x,t)$ for diffusion
on $-\infty<x<\infty$. $C(x,t)$ is a simple over-estimator $C(x,t)\ge c(x,t)$
that can be used to obtain a trial sample for the particle position
$x_{trial}$ on $(0,1)$. Since $C(x,t)$ is a Gaussian, a trial position
can be obtained by scaling and translation of a normally distributed
random number $r_{n}$, $x_{\text{trial}}=(1+r_{n}\sqrt{8t})/2$ {[}$x_{trial}$
can occasionally fall outside $(0,1)$ in which case it is discarded{]}.
With a trial position so selected, the partial sums $S_{m}$ of the
image series for concentration $c(x,t)$ are used as an alternating
sequence of increasingly tight upper and lower bounds convenient for
rejection sampling, as described in appendix A.

At long times, as an approximation for the particle position distribution
it seems reasonable to take the first term of the eigenfunction expansion
{[}c.f. Eq. \ref{eq:c_x_t_long_t}{]}, $\tilde{c}(x,t)=\sin(k\pi x)e^{-\pi^{2}t}$.
Although this function is smaller than the full solution $c(x,t)$
for some $x$, it is still possible to use it to construct a tight
majoring function $C(x,t)\ge c(x,t)$ by multiplying $\tilde{c}$
with a factor $1+g(t)$, so that $C(x,t)=\left[1+g(t)\right]\tilde{c}(x,t)\geq c(x,t)$,
for all $x$. One possible choice for $g(t)$ is

\[
g=\frac{e^{-8\pi^{2}t}}{1-e^{-16\pi^{2}t}}.\]
This particular factor was derived by taking the absolute value of
every term in the eigenfunction series expansion, noting that $|\sin x|\le1$
and that $x^{2}\ge x$ for $x\ge1$, and replacing the square in the
exponential by a linear function. The resulting sum is a geometric
sum and can be evaluated analytically.

Sampling $x_{trial}$ from the majoring distribution is then performed
by picking a uniformly distributed random number $-1\le r<1$, and
setting $x_{trial}=\frac{1}{2}+\frac{1}{2\pi}\arccos r$. In addition
to $x_{trial}$ we also need an estimate for the remainder of the
long-time series. Following a derivation similar to that of $g$,
we find that the function \[
d_{m}=2\frac{e^{-\pi^{2}t(2m+3)^{2}}}{1-e^{-4\pi^{2}t}},\]
bounds the series remainder so that $c_{m}(x,t)-d_{m}\leq c(x,t)\leq c_{m}(x,t)+d_{m}$.
The so defined $C(x,t)$ and $d_{m}$ can be employed for rejection
sampling. To reduce the cost, when it is necessary to compute the
higher order terms of the series expansion for $c(x,t)$, we re-use
the already calculated time exponentials which requires a few multiplications
for each iteration. The time $t_{switch}$ for switchover from the
short time series to the long time series can be selected to optimize
the cost of rejection sampling, similar to the FP propagators described
in the preceding section.

We note in passing that still tighter bounding functions $g$ and
$d_{m}$ can be derived by replacing the infinite sum in $c(x,t)$
with a majoring integral. However the resulting expressions contain
the error function, $\textrm{erf}(x)$, which can be expensive to
numerically evaluate.


\end{document}